\documentstyle[11pt,paspconf,epsfig]{article}
\def\besc{\beta_{\rm esc}}
\def\g{\gamma}
\def\gabs{\gamma_{\rm abs}}

\def\lb{\ell_{\rm B}}
\def\me{m_{\rm e}}

\def\sigmat{\sigma_{\rm T}}

\def\taut{\tau_{\rm T}}
\def\tesc{t_{\rm esc}}
\def\ur{U_{\rm rad}}
\def\ub{U_{\rm B}}

\def\eabs{\nu_{\rm abs}}

\def\a218{\alpha_{2-18}}

\def\aint218{\alpha_{{\rm int},2-18}}
\def\elldiss{\ell_{\rm diss}}

\def\g{\gamma}
\def\Ldiss{L_{\rm diss}}

\def\Ldiss{L_{\rm diss}}

\def\sigmat{\sigma_{\rm T}}
\def\taut{\tau_{\rm T}}
\def\Te{T_{e}}
\def\Tbb{T_{\rm bb}} 

\newcommand{\ltsima} {$\; \buildrel < \over \sim \;$}
\newcommand{\gtsima} {$\; \buildrel > \over \sim \;$}
\newcommand{\lta} {\lower.5ex\hbox{\ltsima}}
\newcommand{\gta} {\lower.5ex\hbox{\gtsima}}

\begin{document}

\title{Thermalization Mechanisms in Compact Sources}

\author{Roland Svensson}
\affil{Stockholm Observatory, SE-133 36 Saltsj\"obaden, Sweden}

\begin{abstract}
There is strong observational evidence that a quasi-thermal population of
electrons (or pairs) exists in compact X-ray sources. It is, however, unclear what
mechanism thermalizes the particles. Here,
two processes, Coulomb scattering and synchrotron self-absorption, that may be
responsible for the thermalization, are reviewed. The parameter
spaces in which respective process dominates are given.

While the Coulomb thermalization mechanism is well-known, this is not the
case for the synchrotron self-absorption thermalization.
We give the arguments that synchrotron self-absorption must act as a 
thermalizing
mechanism in sufficiently compact sources. The emitting and absorbing
electrons then exchange energy efficiently with the self-absorbed 
synchrotron radiation field and are driven towards a {\it relativistic}
or {\it mildly relativistic} thermal distribution in a few synchrotron cooling
times (the ``synchrotron boiler'').
\end{abstract}



\section{Introduction}

Observations with hard X-ray/$\gamma$-ray satellites  such as 
{\it CGRO} OSSE, {\it RXTE}, and {\it BeppoSAX} indicate that
the X/$\gamma$ spectra cut off at a few hundred keV for the majority of active
galactic nuclei (see Zdziarski, this volume; Matt, this volume) and for the
hard states of galactic black hole candidates (see Zdziarski, this volume;
Grove, this volume). It is generally believed that Comptonization
by a quasi-thermal population of electrons (or pairs) is responsible for the 
formation of the X/$\gamma$ spectra from these sources.
In the spectral modeling codes (e.g., Poutanen \& Svensson 1996),
it is normally assumed that the Comptonizing particles have a
Maxwellian distribution. It has been pointed out several times that 
thermalization by Coulomb scattering may not be fast enough as compared to
various cooling mechanisms (such as Compton cooling) and that the
particle distribution therefore will differ from a Maxwellian (e.g.,
Dermer \& Liang 1989; Fabian 1994). On the other hand, it has been noticed that
another thermalization mechanism, synchrotron self-absorption, may operate in
compact plasmas (Ghisellini, Guilbert, \& Svensson 1988;  Ghisellini \&
Svensson 1989).

Here, we review the physics of these two thermalization mechanisms, and 
explore in which contexts each of them may operate. 

\section{Thermalization by Coulomb Scattering}

The approximate time scale, $t_{\rm C}$, for thermalization by Coulomb
(M{\o}ller) scattering
between electrons in nonrelativistic plasmas has been known since
long  (e.g., Spitzer 1956; see Stepney 1983 for relativistic corrections):
\begin{equation}
t_{\rm C} = 4 \frac{t_{\rm T}} {\ln \Lambda}
\Theta^{3/2} (\pi^{1/2} - 1.2 \Theta^{1/4} + 2 \Theta^{1/2}),
\end{equation}
where $t_{\rm T} = (n_e\sigma_{\rm T} c)^{-1}$ is the Thomson time, $n_e$ is the
electron density, $\ln \Lambda \approx 10-20$ is the Coulomb logarithm, and 
$\Theta \equiv kT_e/m_e c^2$ is the dimensionless temperature.
Only with the advent of X-ray astronomy emerged the
understanding that the electrons may reach mildly relativistic temperatures
of 10$^9$ K or more, and that electron-positron pairs may be created.
In such conditions,   
electron-positron (Bhabha) scattering also  becomes important. Here, the
term ``Coulomb scattering'' will be used for all types of ``Coulomb'' interactions.

\subsection{The Relaxation Process}
 
Detailed numerical simulations of the relaxation
and thermalization process were performed by Dermer \& Liang (1989).
As small angle scatterings normally dominate and the fractional energy change
per scattering is small, a Fokker-Planck approach is appropriate.
Dermer \& Liang evaluated the energy exchange and diffusion coefficients
assuming the plasma to have a Maxwellian distribution. Using these 
coefficients in the
simulations is equivalent to studying the relaxation of a test particle
distribution in a Maxwellian background plasma. Nayakshin \& Melia (1998) relaxed
the Maxwellian assumption and computed self-consistent Fokker-Planck coefficients
using the real particle distributions. A Monte Carlo approach was taken
by Pilla \& Shaham (1997) who treated the time evolution of both the pair and
photon distributions in an infinite system  and  who besides Coulomb interactions,
bremsstrahlung, and Compton scatterings also included pair production and
annihilation.

Figure 1 (from Dermer \& Liang 1989) shows the relaxation of a test electron
distribution in a background thermal electron plasma of temperature 511 keV, or,
equivalently, $\Theta$ = 1. The test
electrons had initially a Gaussian distribution centered at 1 MeV and a FWHM of
0.28 MeV. The diffusion process dominates initially and broadens the 
electron distribution that first relaxes at lower energies and only later the
Maxwellian tail forms. Figure 7 in Nayakshin \& Melia (1998) shows the
similar process but here {\it all} electrons are initially Gaussian
(i.e., there is no Maxwellian background).
If the initial distribution is broader than a Maxwellian then the
energy exchange coefficient dominates initially and the low energy electrons gain
energy while the higher energy electrons loose energy thereby narrowing the
distribution towards a Maxwellian (see fig. 6 in Nayakshin \& Melia 1998).

\begin{figure}[t]
\centerline{\epsfig{file=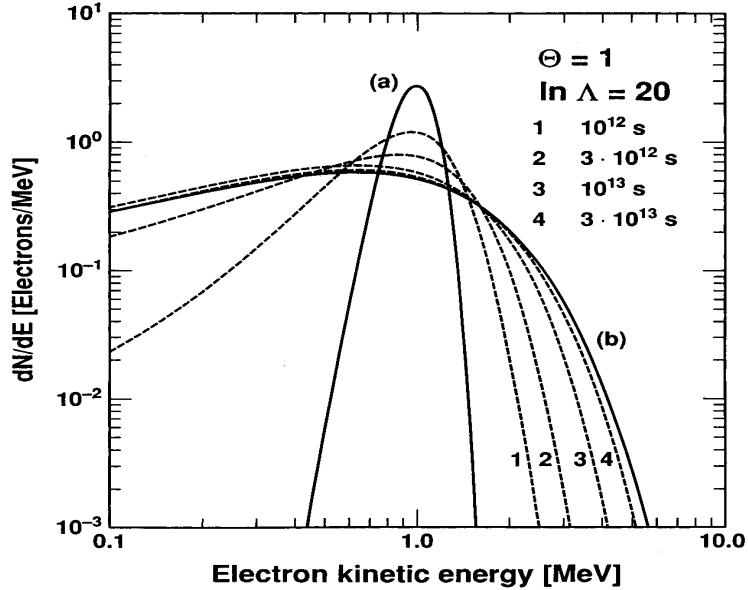,width=10.5cm,height=8cm}}
\caption{Thermalization through electron-electron scattering of an initially
Gaussian electron test distribution (curve a) centered at 1 MeV and a FWHM of 0.28
MeV in a background thermal electron plasma having a Maxwellian distribution of
temperature 511 keV, i.e., $\Theta =1$ (curve b). The dashed curves show the
relaxing distribution at different times, assuming an electron background density
of 1 cm$^{-3}$. The thermalization time scale from eq. (1) is $ 2.6 \times
10^{13}$ s. From Dermer  \& Liang (1989).}
\label{electronrelaxation}
\end{figure}

\subsection{Influence of Cooling Processes on the Steady Electron Distribution}

With increasing temperature, the Coulomb energy exchange rate decreases. Various
cooling processes such as bremsstrahlung, Compton cooling and synchrotron cooling
increases with temperature and eventually the thermalization process
will be inhibited by the cooling first noticeable as a truncation of the
Maxwellian tail. Stepney (1983) noticed that bremsstrahlung cooling will 
prevent thermalization for temperatures larger than about 
5 $\times 10^{10}$ K, and Baring (1987) performed further analysis for
additional cooling processes, as did Ghisellini, Haardt, \& Fabian
(1993).

Including Compton and synchrotron losses in the Fokker-Planck equation
allows for the determination of the steady  distribution function under
the influence of these cooling processes.
Results from Dermer \& Liang (1989) are shown in Figure~2.
It is seen that, for increasing energy densities of radiation and
magnetic fields, the high energy tail of the electron distribution
becomes increasingly truncated and the effective temperature of the
distribution becomes smaller.

What are then the conditions for losses to dominate over Coulomb
thermalization (see, e.g., Fabian 1994)?
The nonrelativistic cooling time scale can be written as 
(see Coppi, this volume) 
\begin{equation}
t_{\rm cool} \approx R/[c \lb (1 + U_{\rm rad} / U_{\rm B})] ,
\end{equation}
where $R$ is the size of the region, $\lb$ is the magnetic compactness
defined in equation (4) below, and $U_{\rm rad}$, $U_{\rm B}$ are the energy
densities in radiation and magnetic fields, respectively.
Comparing with equation (1), one finds that Coulomb scattering cannot maintain
a Maxwellian when
\begin{equation}
\lb (1 + U_{\rm rad} / U_{\rm B}) > \frac  
{\taut \ln \Lambda} {4\Theta^{3/2}},
\end{equation}
where $\taut = n_e \sigmat R$ is the Thomson depth.

\begin{figure}[t]
\centerline{\epsfig{file=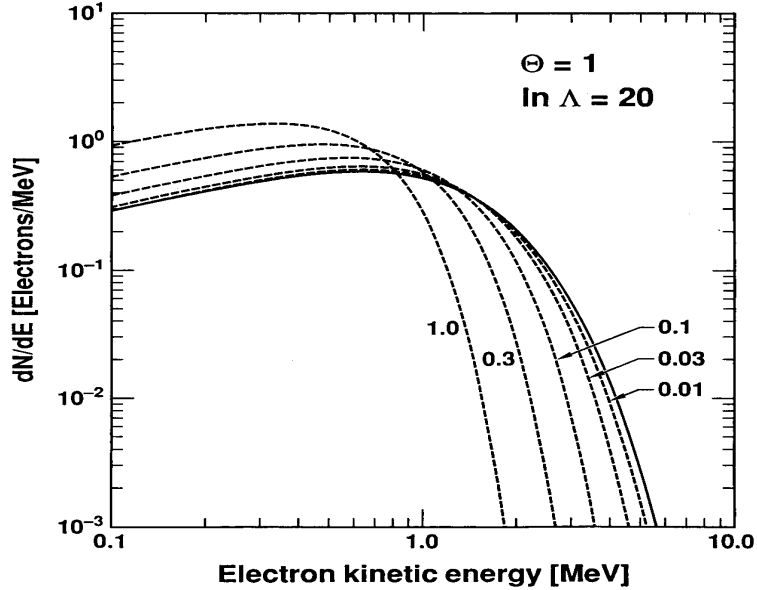,width=10.5cm,height=8cm}}
\caption{
Truncation of a thermal electron distribution with temperature 511 keV due to
Compton and synchrotron losses. Solid curve shows a Maxwellian of
temperature 511 keV. Dashed curves show the relaxed distributions at different
values of the parameter $(U_{\rm rad} + U_{\rm B})/ n_e$ in units of MeV.
From Dermer  \& Liang (1989).}
\label{truncation}
\end{figure}

\subsection{Can Coulomb Collisions Thermalize Pair Coronae and
Active Pair Regions?}

Two different numerical codes (Stern et al.\ 1995a; Poutanen \& Svensson 1996) have
been used to study radiative transfer and Comptonization in pure pair coronae in
energy and pair balance (see, e.g., Svensson 1997 for a review). For coronae of a
given geometry and in energy 
balance, there exists a unique $T_e - \taut$ relation, 
where $\Te$ is the volume-averaged coronal temperature 
and $\taut$ is a characteristic 
Thomson scattering optical depth of the coronal region.
In Figure 3a,
this relation is shown for different geometries.
The results for active regions are connected by {\it dotted} curves.
For comparison we also show the slab results from Stern et al.\ (1995b)
using an iterative scattering method code ({\it dashed curve}).

Solving the pair balance for the obtained combinations of ($\Theta$, $\taut$)
gives a unique dissipation compactness, $\elldiss$ (see Ghisellini \& Haardt 1994
for a discussion). Here, the local dissipation compactness,
$\elldiss \equiv ( \Ldiss /h ) (\sigmat /  m_e c^3)$,
characterizes the dissipation with $\Ldiss$ being the power providing uniform
heating in a cubic volume of size $h$ in the case of a slab of height $h$, 
or in the whole
volume in the case of an active region of size $h$.
Figure 3b
shows the $\Theta$ vs. $\elldiss$ relations.

\begin{figure}[h]
\centerline{\epsfig{file=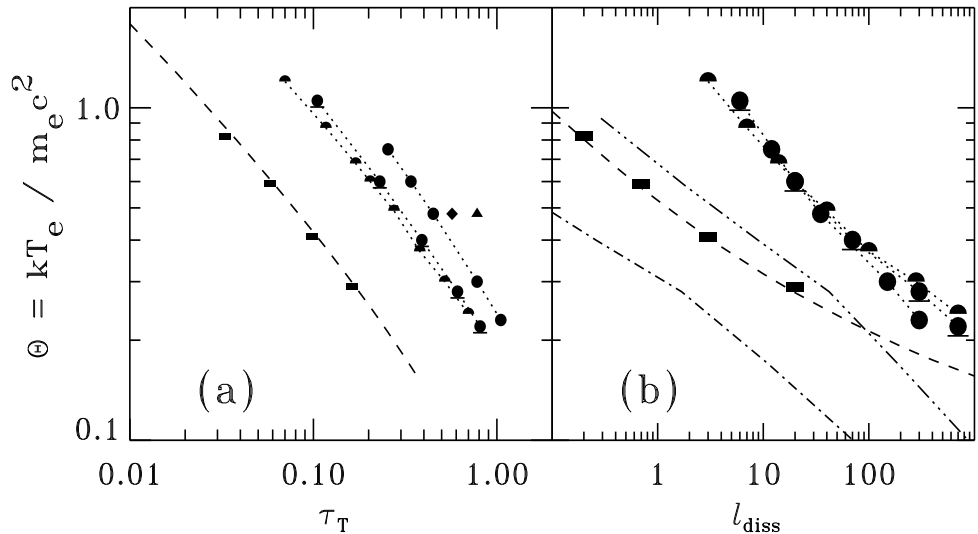,width=13.5cm,height=8cm}}
\caption{\small 
Dimensionless volume-averaged temperature, $\Theta \equiv kT_e/ m_e c^2$,
vs. Thomson scattering optical depth, $\taut$, in panel (a), and vs.
 dissipation compactness, $\elldiss \equiv ( \Ldiss /h )$ 
$(\sigmat /  m_e c^3)$ in panel (b), for
a steady X-ray emitting  plasma region in pair and energy
balance on or above a cold disk surface of black body temperature,  $k\Tbb$ = 5 eV.
The plasma Compton scatters reprocessed soft black body photons from
the cold disk surface. 
{\it Solid rectangles} and {\it dashed curve} show results
from a nonlinear Monte Carlo code (Stern et al.\ 1995a) and  an iterative scattering
method  code (Poutanen, \& Svensson 1996), respectively, for the case of a
plane-parallel slab corona.   Results using the Monte Carlo code for individual
active pair regions are shown for
{\it hemispheres} located on the disk surface;
surface spheres also located on the surface ({\it underlined spheres});
spheres located at a height of 0.5$h$ ({\it spheres}),
1$h$ ({\it diamond}), and 2$h$ ({\it triangle}), where $h$ is the radius
of the sphere.
The results for each type of active region are connected by
{\it dotted} curves.  
The {\it dash-dotted} and {\it dash-dot-dot-dotted} curves  in panel (b)
show the critical compactness as function of $\Theta$ above which 
thermalization by M\o ller and Bhabha scattering is not achieved for the
cases of pair slabs and surface spheres, respectively.
(Unpublished results by Stern et al.; see also Stern et al. 1995b; Svensson 1997).
}
\label{temptaut}
\end{figure}

The question arises whether the electrons can thermalize or not
for the conditions, $\Theta$, $\taut$, and $\elldiss$, in 
Figure 3.
Energy exchange and thermalization
through M\o ller ($e^{\pm}e^{\pm}$)
and Bhabha ($e^+e^-$) scattering compete with various loss mechanisms,
with Compton losses being the most important for our conditions.
The thermalization is slowest and the Compton losses largest for the
higher energy particles in the Maxwellian tail.
Instead of using the approximate equation (3),
we use the detailed simulations
by Dermer \& Liang (1989, their fig. 8) to find the critical compactness 
above which the deviation of the electron distribution 
at the Maxwellian mean energy is more than a factor
$e
\approx 2.7$.  The {\it dash-dotted} and {\it dash-dot-dot-dotted} curves in
Figure~\ref{temptaut}b 
show this critical compactness for slabs and for surface spheres, respectively.
In agreement with Ghisellini, Haardt, \& Fabian (1993), we find that M\o ller  
and Bhabha scattering cannot compete with Compton losses in our pair
slabs and active regions. 

The problem then arises of what mechanism can thermalize the apparently
thermal electron distribution in compact sources. One such mechanism
is cyclo/synchrotron absorption.

\section{Thermalization by Cyclo/Synchrotron Absorption}

\subsection{A Brief History of Synchrotron Thermalization}

Ever since the classical interpretation by Shklovskii in the 1950s
of the radiation
from the Crab nebula as being synchrotron radiation, this process has played an
important role in our interpretation of the non-thermal radiation from
a wide variety of astronomical objects.
In general, the electron distribution has been assumed to be a
power law or nearly a power law.
Much less attention has been paid to what happens to the
electron distribution at self-absorbed electron energies.
 
The theory for synchrotron radiation was developed in the 1950s
(see, e.g., reviews by Ginzburg and Syrovatskii 1965, 1969; Pacholczyk 1970).
The emission
and absorption coefficients for single relativistic electrons
as well as for ensembles of relativistic electrons having power law or
thermal distributions were calculated in the  1950s and 1960s.
For power law distributions, the absorption coefficient increases towards lower
photon energies. Below some photon energy, $\eabs$, the source becomes optically
thick to synchrotron self-absorption resulting in an intensity proportional to
$\nu^{5/2}$, obtained from the ratio of
emission and absorption coefficients (Le Roux 1961).
A finite $\eabs$, of course, requires that the source is finite. Below we
call this  and its consequences "finite source effects".  

This early work was, in general, applied to extended sources
where the cooling time at self-absorbed particle energies is
longer than other relevant time scales
(such as the age of the source or the dynamical time scales).
It was therefore natural to assume that the self-absorbing electron
distribution below the Lorentz factor, $\gabs$, of the electron
(emitting at the frequency $\eabs$ where the source becomes optically thick) is
unaffected by self-absorption and simply maintains the power law distribution of
the injected electrons.

In the late 1960s, it became increasingly clear that, in compact sources
or on long time scales, the self-absorbed electron distribution
$N(\g)$ will evolve under the influence of synchrotron emission
and absorption. What are then the possible equilibrium solutions
at self-absorbing Lorentz factors towards which $N(\g)$ would relax?
In the important papers by Rees (1967) and McCray(1969), it was shown
that power law distributions, $N(\g) \propto \g^{-s}$
with $s = 2$ and 3, are equilibrium solutions to the kinetic
equations.
Rees (1967), however, also found that the solution with $s=3$
is unstable and would evolve away from $s =3$ if slightly perturbed.
McCray (1969) showed this explicitly by numerically calculating
the time dependent evolution of initial power law distributions
in an infinite source.
Rees predicted and McCray confirmed that the high energy electrons in a flat
($s<3$) initial power law would tend to evolve into
a quasi-Maxwellian distribution.
McCray (1969), furthermore, emphasized the importance
of finite source effects on the evolution.
In particular, for power laws with $s<3$, the self-absorbing
electrons would gain energy absorbing slightly more energy
than they emit, while the electrons radiating in the optically thin limit
lose energy by radiating much more energy than they absorb.
All electrons would therefore tend to gather at $\gabs$ developing a peak
there (as was already emphasized by Rees 1967).

It must be emphasized that the relaxation of self-absorbing electrons
takes place through the energy exchange with the radiation field,
which in its turn is determined by the particle distribution. This is the 
``synchrotron boiler'', a terminology coined by Ghisellini, Guilbert, \& 
Svensson (1988). 

\subsection{Rise and Fall of yet another Paradigm}

In a series of papers in the 1970s, Norman and coworkers further
developed the concept {\it Plasma Turbulent Reactor} (PTR)
introduced by Kaplan and Tsytovich (1973).
Originally the turbulence feeding the electrons was thought to be plasmons.
In practice, however, the PTR is exactly the self-absorbing synchrotron
source considered here, as photons are the only plasma modes with
sufficiently small damping rate that they can mediate energy transfer
from one electron to another (Norman 1977).
Norman and ter Haar (1975) and Norman (1977) essentially repeated the
analysis of McCray (1969) using quite different notation
and definitions but arriving at the same conclusions that
$N(\g) \propto \g^{-2}$ and $N(\g) \propto \g^{-3}$ are the only
steady power law equilibrium solutions.
It is important that they noted that the $N(\g) \propto \g^{-2}$
solution corresponds to a finite electron flux upwards along the
energy axis, while $N(\g) \propto \g^{-3}$ corresponds to zero
electron flux.
They argued that $N(\g) \propto \g^{-3}$ was the most physical solution
as the synchrotron time scales establishing this distribution are
shorter than other time scales.
Although being aware of possible finite source effects,
they considered them not to influence the electron distribution
at Lorentz factors $\ll \gabs$.

The self-absorbed solution, $N(\g) \propto \g^{-3}$, was considered
sufficiently important in explaining power law spectra from a variety of
sources that Norman and ter Haar (1975) called the PTR a
{\it new astrophysical paradigm}.
Norman and coworkers, however, do not seem
to have considered the stability of the $N(\g) \propto \g^{-3}$
solution.

The work of Rees (1967) and McCray (1969) indicates that a
Maxwellian distribution may be the only {\it stable}
equilibrium solution. This was, however, not rigorously established causing
Ghisellini, Guilbert \& Svensson (1988, GGS88)
to numerically determine the
steady solutions of the kinetic equations including
physical boundary conditions (i.e., correct Fokker-Planck coefficients
at subrelativistic energies, and the accounting for finite source effects
at large energies).
As $\gabs$ typically is of the order 10-100 in compact radio sources and the
development of the self-absorbed distribution takes place at mildly relativistic
energies, they used expressions and equations valid at any energy.
Furthermore, in order to obtain steady solutions the particle injection
had to be balanced by a sink term (escape or reacceleration).
Injecting a power law proportional to $\g^{-3}$
(i.e., with the equilibrium slope) GGS88 found that the steady solution
was a {\it Maxwellian} with a temperature corresponding to the mean energy
of the injected electrons.
Similarly, an injected power law proportional to $\g^2$
(essentially corresponding to monoenergetic injection at some large
Lorentz factor $\gg \gabs$) led to the establishment of a Maxwellian
just below $\gabs$. The injected electrons cool
until reaching $\gabs$ where they thermalize exchanging energy
with the self-absorbed radiation field.
Additional studies were made by de Kool, Begelman, \& Sikora (1989)  and
Coppi (1990). 
With these works it appears that the PTR paradigm of Norman and ter Haar (1975)
has been shown to be invalid.

\subsection{Relaxation by Cyclo/Synchrotron Absorption} 

The works so far that explicitly have {\it demonstrated}
the formation of a Maxwellian through synchrotron self-absorption
are the numerical simulations of GGS88, Coppi (1990), and 
Ghisellini, Haardt, \& Svensson (1998).
Here, we review some of the results in the last paper.
In the numerical simulations, a kinetic equation for
the electron distribution is solved.
The kinetic equation, which also in this case takes the form of a Fokker-Planck
equation (see derivation in McCray 1969), includes Compton and synchrotron cooling,
synchrotron absorption (heating), electron injection, and electron
escape. Even though various source geometries are discussed, the radiation field 
is assumed to be given by the steady slab solution, which is correct to the order
of unity.
 
The simulations  consider a region of size $R$ with a
magnetic field of  strength $B$,
into which some distribution of electrons are injected with a
power $L$. In a steady state, this power 
emerges either as radiation or as the power of escaping electrons.
The electrons are assumed to escape at the speed
$v_{\rm esc}=c \besc ={R/ \tesc}\,$
where $\tesc$ is the escape time. 
Convenient parameters describing compact sources are the
injection compactness,
$\ell_{\rm inj}$,
and the magnetic compactness, $\lb$, defined as
\begin{equation}
\ell_{\rm inj}={L \over R} {\sigmat \over \me c^3}\,; \qquad
\lb= { \sigmat  \over \me c^2} R U_{\rm B}\,,
\end{equation}
where $\sigmat$ is the Thomson cross section and
$\ub= B^2/8 \pi$ is the magnetic field energy density.
Note that ${\ur / \ub} \approx (9 / 16 \pi) (\ell_{\rm inj}/ \lb) (1+\taut)$,
where the numerical factor is dependent on the source geometry. 
The problem we consider has the following parameters:
$\ell_{\rm inj}$, $\lb$, $\besc$, and either $R$ or $B$.
Further parameters are those describing the shape of the injected electron
distribution.

For steady state electrons emitting and absorbing synchrotron photons, the
cooling (emission) and absorption/diffusion time scales are balanced and thus
equal. The synchrotron cooling time scale can thus be taken to be the
thermalization time scale.
From Equation (2), it is then clear that self-absorbing electrons will 
thermalize before they escape when $\lb \gta 1$. The simulations
shown below have  $\lb$ = 10 and 30.

Figure 4 shows the relaxation due to cyclo/synchrotron absorption of an electron
distribution towards the  equilibrium Maxwellian distribution. The injected
electrons have a Gaussian  energy distribution peaking at $\gamma=10$. 
Each curve is labeled by the time in units of $R/c$. The shape of the
equilibrium distribution is reached in about
$\sim 0.1 R/c$, about equal to the cyclo/synchrotron cooling time.
With the assumed input parameters, the
synchrotron terms (emission,  absorption and energy diffusion) in the
kinetic equation are dominant over  Compton losses. Gains and losses in
this case almost perfectly balance.  As a result the equilibrium electron
distribution is a Maxwellian. Figure 4  also shows that the high energy
part of the Maxwellian distribution is formed  earlier than the low
energy part, due to the higher efficiency of  exchanging photons of the
high energy electrons. A slower evolution takes place after $0.1(R/c)$,
as the balance between electron injection and electron escape is
achieved on a time scale of a few $t_{\rm esc}$. Only then have
both the shape and the amplitude of the electron distribution
reached their equilibrium values.

\begin{figure}[t]
\centerline{\epsfig{file=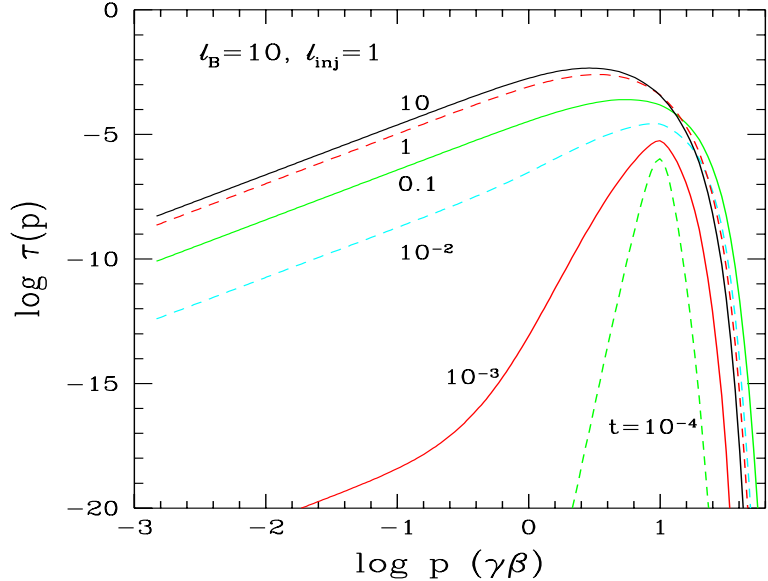,width=11cm,height=8cm}}
\caption{Electron distribution, $\tau(p) \equiv \sigmat R N(p)$, evolving due to
cyclo/synchrotron emission/absorption/diffusion. 
Curves are labeled by times  in units of
$R/c$. Parameters are $\ell_{\rm inj}=1$,  $\ell_{\rm B}=10$, $R=10^{13}$ cm
(or $B\simeq 5.5\times 10^3$ G),  and $\besc = 1$. 
The injected distribution is a Gaussian centered at $\gamma=10$.
From Ghisellini, Haardt, \& Svensson  (1998).}
\end{figure}

\subsection{Influence of Cooling Processes on the Steady Electron Distribution}
 
The equilibrium distributions for different values of the injected 
compactness are shown in Figure 5. The magnetic compactness is set to 
$\ell_{\rm B}=30$, 
corresponding to $B=10^4$ G for $R=10^{13}$ cm (from Eq. 4). 
In all cases, the injected distribution is a peaked function with an exponential
high energy cut-off.
The  mean injected Lorentz factor is $<\gamma>\simeq 5$ and essentially
all electrons are below $\gabs$. 
It is apparent from Figure 5 that the electron distribution is a
quasi-Maxwellian at all energies as long as
$\ell_{\rm inj} \ll \ell_{\rm B}$.
This is a consequence of an almost perfect balance between
synchrotron gains (absorption) and losses, while Compton losses are
only a small perturbation. As $\ell_{\rm inj}$ increases towards $\ell_{\rm B}$, 
Compton losses  become increasingly important, competing with the
synchrotron processes.  At high energies, losses overcome gains, and the electrons
diffuse downwards  in energy, until subrelativistic energies are reached. 
In this energy regime, the increased efficiency of synchrotron gains 
(relative to losses) halts the systematic downward diffusion in energy, 
and a Maxwellian can form (see Ghisellini \& Svensson 1989). 
The temperature of this part of $N(\gamma)$
can be obtained by fitting a Maxwellian to the the 
low energy part of the distribution, up to energies just above the peak of
the electron distribution.

\begin{figure}[p=t]
\centerline{\epsfig{file=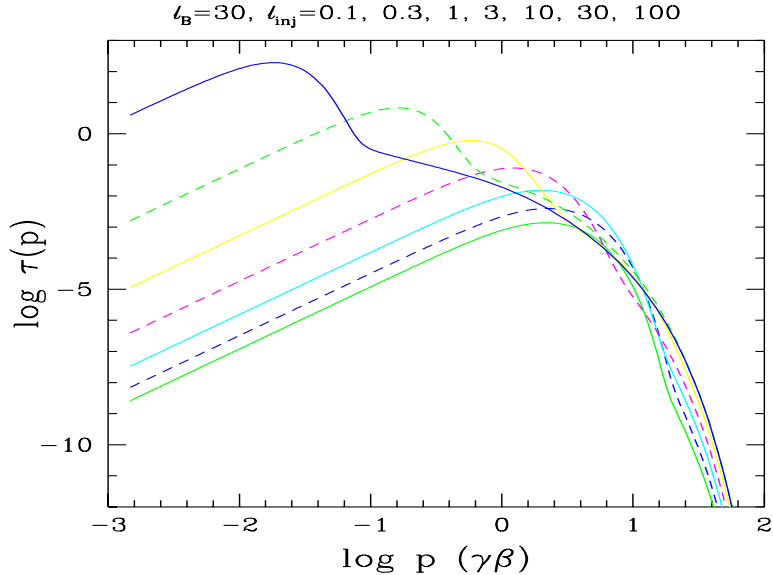,width=11cm,height=8cm}}
\caption{Steady equilibrium electron distributions due to cyclo/synchrotron
emission/absorption/diffusion and Compton cooling
for different injected compactnesses (decreasing from top to
bottom). Further parameters are $\ell_{\rm B}=30$, $R=10^{13}$ cm, 
$\besc = 1$. The corresponding magnetic field is $B=  10^4$ G.
The mean
injected Lorenz factor is about 5. Increasing $\ell_{\rm inj}$ implies increased
Compton cooling resulting in a shift of the quasi-Maxwellian towards lower
temperatures. From Ghisellini, Haardt, \& Svensson  (1998).}
\end{figure}

The resulting temperatures are plotted in Figure 6 as a function of 
$\ell_{\rm inj}$.
For  $\ell_{\rm inj} \lta 1$, the temperature
is approximately constant, while it  decreases for $\ell_{\rm inj} \gta 1$. 
From Equation (3) (with $U_{\rm rad} / U_{\rm B}$ set to zero), we see that
thermalization by synchrotron self-absorption dominates when 
$\ell_{\rm B} >$ $ \tau_{\rm T} \ln \Lambda / 4 \Theta^{3/2}$ 
(assuming $\Theta \lta 1$). 
The Coulomb process thus dominates
for small temperatures and large $\tau_{\rm T}$
(i.e., large electron densities). We need to know 
$\taut$ for our simulations. The balance between electron
injection and escape in our model gives a Thomson optical depth of
$\tau_{\rm T} = (3 / 4 \pi)  
(\ell_{\rm inj} / \besc <\gamma>).$
For the simulations in Figure 5, the optical depth increases from 
$\tau_{\rm T}=5\times 10^{-3}$ 
for $\ell_{\rm inj}=0.1$ to $\tau_{\rm T}=5$ for $\ell_{\rm inj}=100$.
Using the expression for $\taut$, we find that thermalization by synchrotron
self-absorption then dominates over Coulomb scattering for temperatures  
$ \Theta > 0.11 (\ln \Lambda / <\gamma>)^{2/3}  
(\ell_{\rm inj} / \ell_{\rm B})^{2/3} $.  
For the parameters of the simulations in
Figures 5 and 6
 and $\ln \Lambda = 20$,
the condition  becomes $\Theta > 0.03 (\ell_{\rm inj})^{2/3}$, which is plotted as
the solid line in Figure 3. One sees that  that synchrotron
self-absorption dominates the thermalization for
all cases with $\ell_{\rm inj}$ smaller than about 10 . For the cases
$\ell_{\rm inj} =$ 30 and 100, one cannot neglect Coulomb
thermalization.  
 
\begin{figure}[t]
\centerline{\epsfig{file=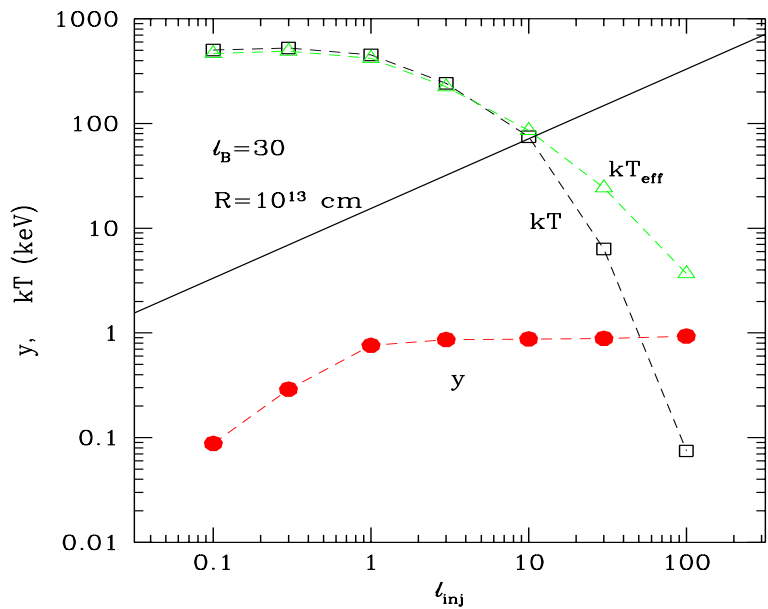,width=11cm,height=8.5cm}}
\caption{Temperatures of the quasi-Maxwellian part of steady equilibrium
electron distributions shown in Fig. 5. Above the solid
line, synchrotron self-absorption dominates over Coulomb
exchange as the thermalization mechanism. The solid dots show the
Comptonization $y$-parameter. From
Ghisellini, Haardt, \& Svensson  (1998). }
\end{figure}

\subsection{Spectra from Steady Electron Distributions}

In Figure 7, the radiation spectra corresponding to four of the 
equilibrium electron distributions in Figure 5 are shown. 
Each spectrum consists of several continuum components: 
\begin{itemize} 
\item
a self--absorbed synchrotron spectrum (S);
\item
a Comptonized synchrotron spectrum (SSC);
\item
a reprocessed thermal soft component (bump);
\item
a component from Comptonization of thermal bump 
photons (IC);
\item
a Compton reflection component.
\end{itemize}

\begin{figure}[t]
\centerline{\epsfig{file=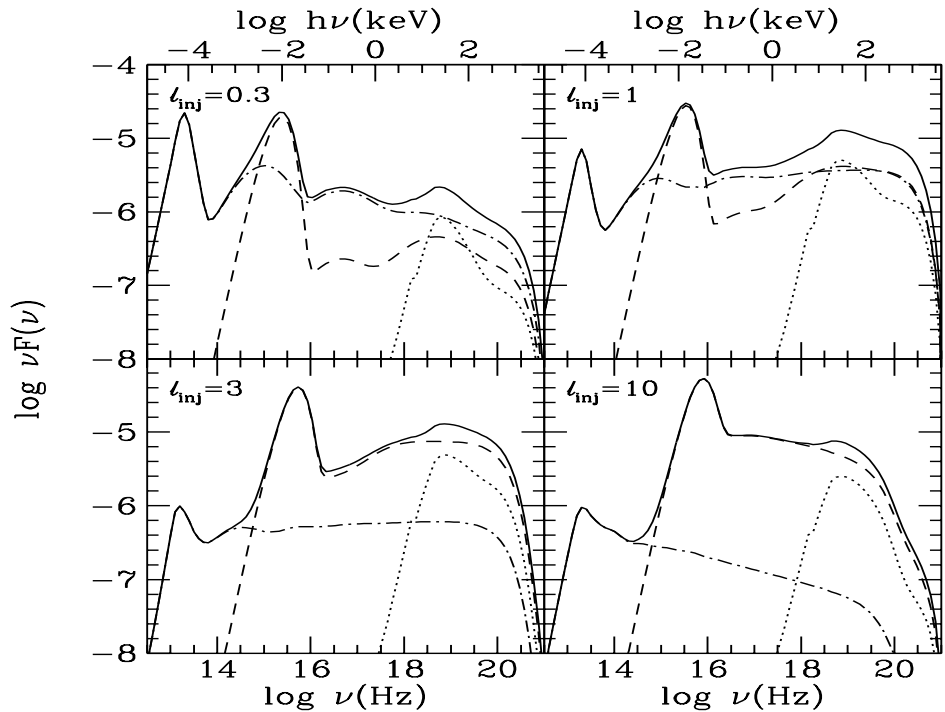,width=13.5cm,height=10cm}}
\caption{Radiation spectra calculated using four of the electron distributions
shown in Fig. 5. The different spectral components are 
the reprocessed thermal bump and IC components (dashed curves), 
the synchrotron and SSC components (dash--dotted curves), and 
the Compton reflection component (dotted curves). The total spectra
are shown by the solid curves. Note that the chosen parameters
($R=10^{13}$ cm, and the corresponding magnetic field $B=  10^4$ G)
correspond to conditions expected near black holes in active galactic nuclei.
From Ghisellini, Haardt, \& Svensson (1998).}
\end{figure}

Details of the spectral calculations are given in
Ghisellini, Haardt, \& Svensson (1998).
Some features in Figure 7 may be noticed.
For $\ell_{\rm inj} < 1$, the Compton $y$-parameter is less than unity
(see Fig. 6) making
the  Compton losses relatively unimportant relative the self-absorbed
synchrotron radiation. The large value of $\Theta$ makes the Comptonized 
spectra bumpy.
The 2--10 keV
band is dominated  by the SSC component, rather than by the IC.
The thermal bump and the X--ray flux are thus not directly related.
This is contrary to the common  interpretation of  
the X-ray emission in Seyfert galaxies as being due to
Comptonization of thermal bump photons.

For $\ell_{\rm inj} > 1$, the Compton cooling dominates and limits
the $y$-parameter to unity. The smooth IC power law dominates over the S and SSC
components.  For $\ell_{\rm inj}\lta 3$, the high energy spectral cut-off
can be  described by an exponential, since the electron distribution 
is a quasi-Maxwellian in the entire energy range.
For $\ell_{\rm inj}\gta 3$, the electron distribution is more complex 
(see Fig. 5), resulting in a more complex spectral cut-off.

The choice of $R=10^{13}$ cm (or $B=  10^4$ G) in Figure 7 corresponds
to the case of active galactic nuclei (AGN). Ghisellini, Haardt, \& Svensson
(1998) also study the case of galactic black holes choosing
$R=10^{7}$ cm (or $B=  10^7$ G) for the same set of $\lb$ and
$\ell_{\rm inj}$. The large magnetic field and small size move the
S and bump peaks to larger frequencies. The main difference is that now
the synchrotron component is not completely self-absorbed leading
to an optically thin synchrotron component from the highest energy electrons.
More soft synchrotron photons enhances the SSC component relative the IC
component as compared to the AGN case.

\section{Final Remarks}

First, we note that
$\lb > 1$ is needed for synchrotron self-absorption to operate
efficiently (from eq. 2). A Maxwellian is then formed
with the same mean energy as the injected electrons, assuming that
self-absorption operates at essentially all electron energies
of interest.
If, furthermore, Compton cooling is important,
i.e., if $\ell_{\rm inj} > \lb$, then the Maxwellian is modified and is 
shifted to lower energies (temperature).

Second, we note that the 
criterion for Coulomb vs synchrotron thermalization
in the case when all electrons are self-absorbing (i.e. radiating
optically thick synchrotron radiation)
is more complex (eq. 3).
Essentially, the same criterion is valid for comparing
Coulomb thermalization in the case when the major part of the
electrons radiate optically thin synchrotron radiation
(or Compton radiation) truncating the Maxwellian.
Which of the two cases that is valid
depends on whether the energy, $(\gabs - 1) m_e c^2$, 
of the electrons radiating at the
photon energy where the absorption optical depth is unity
is much larger or much smaller than $kT_e$.

There should also be a region in
parameter space where both Coulomb and synchrotron
thermalization operates simultaneously (assuming that
most electrons radiate optically thick radiation).
Here, Coulomb thermalization should dominate at lower 
electron energies and
synchrotron thermalization at larger energies.
However, nobody seems so far to have solved the Fokker-Planck
equations to study the thermalization process including both Coulomb
scattering and synchrotron self-absorption.
 
Ultimately, the thermalization process should be put in a
realistic context.
Mahadevan \& Quataert (1997) studied the importance of thermalization 
in advection-dominated flows onto black holes under the conditions
considered in such flows (e.g., close to free-fall, equipartition magnetic
fields). Comparing the thermalization time scales with the accretion time
scale (equivalent to $\tesc$ in our discussion), they
found that thermalization did not occur at large radii and small
accretion rates. However, at sufficiently large accretion
rate, synchrotron thermalization becomes important, and at even
larger rates (and thus larger densities) Coulomb thermalization
starts operating.

Another scenario for generating the X-ray radiation from compact sources
is that of a corona or magnetic flares atop an accretion disks.
The typical condition for the flare regions is that
the magnetic energy density should dominate the radiation energy density,
i.e., that $\lb \gta \ell_{\rm inj} > 1$, which should ensure
that cyclo/synchrotron self-absorption acts as a very efficient
thermalizing mechanism in such regions.

\newpage
 
\acknowledgments

I appreciate a more than decade-long collaboration with G. Ghisellini on the
issues discussed in this review. I thank J. Poutanen and A. Beloborodov
for valuable comments.
This work is supported by the Swedish Natural  Science Research Council and the
Swedish National Space Board.

\end{document}